\documentclass[twocolumn]{aastex62}
\usepackage{amssymb}
\usepackage{amsmath}
\usepackage{array,multirow}
\usepackage{comment}
\usepackage{enumerate}
\usepackage{bm}

\newcommand{\cosmic}{{\texttt{COSMIC}}}

\newcommand{\Porb}{\ifmmode {P_{\rm orb}}\else${P_{\rm orb}}$\fi}

\newcommand{\Msun}{\ifmmode {{M_\odot}}\else{$M_\odot$}\fi}
\newcommand{\Mtot}{\ifmmode {{M_{\rm tot}}}\else{$M_{\rm tot}$}\fi}
\newcommand{\RV}{\ifmmode {{\rm RV}}\else RV \fi}
\newcommand{\bigG}{\ifmmode {\mathcal{G}}\else${\mathcal{G}}$\fi}

\newcommand{\Mi}{M_{\rm He}}
\newcommand{\Mf}{M_{\rm BH}}

\newcommand{\ogle}{MOA-11-191/OGLE-11-0462}

\submitjournal{ApJ}

\shorttitle{Black Holes Kicks}

\begin{document}

\title{{\bf Constraining Black Hole Natal Kicks with Astrometric Microlensing}}

\author[0000-0001-5261-3923]{Jeff J. Andrews}
\affiliation{Center for Interdisciplinary Exploration and Research in Astrophysics (CIERA), 
1800 Sherman Ave., 
Evanston, IL, 60201, USA}
\affiliation{Physics Department, University of Florida, 2001 Museum Rd., Gainesville, FL 32611}
\email{jeffrey.andrews@northwestern.edu}

\author[0000-0001-9236-5469]{Vicky Kalogera}
\affiliation{Center for Interdisciplinary Exploration and Research in Astrophysics (CIERA), 
1800 Sherman Ave., 
Evanston, IL, 60201, USA}
\affiliation{Department of Physics and Astronomy, 
Northwestern University,
2145 Sheridan Rd., 
Evanston, IL 60208, USA}

\begin{abstract}
Multiple pieces of evidence suggest that neutron stars receive large kicks when formed from the remnant of a collapsing star. However, the evidence for whether black holes (BH) receive natal kicks is less clear, reliant on weak constraints from the analysis of BH X-ray binaries and massive runaway and walkaway stars. Here we show for the first time that recent microlensing detections offer a new method for measuring the kicks BHs receive at birth. When a BH is identified through both photometric and astrometric microlensing and when the lensed star has a known distance and proper motion, the mass, distance and proper motion of the BH can be determined. We study the runaway velocities for components of eccentric binaries disrupted during a supernova, finding the peculiar velocity correlates strongly with the kick a BH received a birth, typically within 20\%, even when the natal kick is smaller than the orbital velocity. Therefore, by measuring the peculiar velocity of a BH or other compact object that formed from a binary that disrupted during core collapse, we are in effect measuring the natal kick that object received. We focus on MOA-2011-BLG-191/OGLE-2011-BLG-0462, an isolated, single BH detected by microlensing and consider a range of possible formation scenarios, including its formation from the disruption of a binary during a supernova event. We determine that MOA-2011-BLG-191/OGLE-2011-BLG-0462 has a Milky Way orbit consistent with a thick disk population, but if it were formed within the kinematic thin disk it received a natal kick $\lesssim$100 km s$^{-1}$. 
\end{abstract}


\section{Introduction}
\label{S:intro}

As some of the most energetic phenomena in the Universe, supernovae (SN) release $\sim$10$^{51}$ erg s$^{-1}$ into the surrounding environment in seconds \citep{woosley1986}; even slight asymmetries can have significant consequences for the kinematics of the neutron star (NS) and black hole (BH) remnants left behind \citep{shklovskii1970, janka2013, wongwathanarat2013}. By measuring the proper motions of a sample of pulsars, \citet{lyne1994} first and then later \citet{hobbs2005} determined that NSs receive kicks at birth of several hundred km s$^{-1}$. Careful analysis of pulsar binary systems largely confirm the requirement for NSs to be born with large kick velocities \citep[see][and references therein]{wong2010, tauris2017}. 

It is not clear whether BHs are born with similarly large kick velocities. Current constraints on BH natal kicks rely on a variety of data sets. The most stringent limits are placed when analyzing individual X-ray binaries (XRBs) known to host BH accretors \citep{mirabel2001, jonker2004, willems2005, fragos2009, repetto2012, wong2012,  wong2014, repetto2015, repetto2017, mandel2016}. At different levels of sophistication, these studies use a combination of binary stellar evolution models and Milky Way orbital modelling to uncover the formation of individual systems, thereby placing limits on the velocity of BH natal kicks. Although the firmness of the constraints differ for each modelled system, it appears that at least one XRB provides strong evidence that some BHs receive natal kicks  \citep[e.g., $>$80 km s$^{-1}$ for XTE J1118$+$480;][]{fragos2009}.

At the same time, at least two ``failed'' SN candidates have been identified \citep{kochanek2008, gerke2015, reynolds2015} in which a massive star disappeared, implying these stars collapsed into a BH but emitting either very little or none of the radiation characteristic of a SN \citep{smartt2009}. To produce a failed SN, a massive star must undergo a prompt collapse, likely without developing the asymmetries required to impart large kicks, a possibility supported by some SN models \citep{lovegrove2013, sukhbold2016, patton2020}. Indeed, \citet{mirabel2003} used a combination of orbital dynamics, proper motion measurements, and a lack of detected supernova remnant to suggest that the well-studied XRB Cygnus X-1 may have formed through such a mechanism.

Recently, \citet{callister2021} and \citet{stevenson2022} have used the catalog of merging binary BHs detected by gravitational-wave observatories \citep{gwtc2, gwtc3} to show that, if the population is formed through isolated binary evolution then BHs must receive significant natal kick velocities ($\sim$100 km s$^{-1}$) in order to explain the observed spin and spin-orbit misalignment angle distributions. 

Finally, a number of studies have focused on a subset of massive stars that move with abnormally large velocities, exploring the possibility that these runaway and walkaway stars are the former binary companions to collapsing stars \citep{blaauw1961, van_rensbergen1996, de_donder1997, dray2005, eldridge2011, boubert2018, renzo2019, aghakhanloo2022}. In this scenario, the mass loss from the collapsing star and possible natal kick imparted to the newly formed compact object disrupted the binary, leaving behind the unbound companion. \citet{renzo2019} in particular focus on the effects that BH kicks have on the masses and kinematics of such unbound OB stars. Although the effects of BH kicks on massive star populations can be significant, placing constraints using observed populations is challenging; observational selection effects are significant and any constraint is population dependent, requiring an accurate knowledge of massive binary star population characteristics at birth. 

Recently, a new observational data set affords the opportunity to measure the kicks that BHs receive at birth: BHs detected through microlensing. Several free-floating BH candidates have been identified through the long-timescale magnification they induce on background stars \citep{bennett2002, mao2002, wyrzykowski2016, wyrzykowski2020}; however these have all been measured photometrically only. Recently, \citet{sahu2022} and \citet{lam2022} have both separately presented six years of data, both photometric and astrometric, of MOA-2011-BLG-191/OGLE-2011-BLG-0462 (hereafter MOA-11-191/OGLE-11-0462). The astrometric data in particular breaks degeneracies in the model parameters and allows for the measurement of the lens's mass, distance, and transverse velocity. Using the same data, the groups find different solutions for the lens properties. \citet{sahu2022} report a 7.1$\pm$1.3\,$\Msun$ BH at a distance of 1.58$\pm$\,0.18 kpc, moving with a tangential velocity of $\simeq$45\,km s$^{-1}$. \citet{lam2022} find MOA-11-191/OGLE-11-0462 is either a 3.7\,$\Msun$ BH or a 2.1\,$\Msun$ NS or BH, depending on the relative weighting between the photometric and astrometric data. Both \citet{sahu2022} and \citet{lam2022} hint that the observed tangential velocity could be due to a kick the BH received at birth. Here we aim to explore this possibility, with the goal of inferring new constraints on BH natal kicks. In Section~\ref{sec:method} we provide our method for calculating the runaway velocities of stars in eccentric binaries, while we describe the results of those calculations in Section~\ref{sec:results}. We consider possible formation scenarios of MOA-11-191/OGLE-11-0462 in Section~\ref{sec:discussion}, and provide some concluding thoughts in Section~\ref{sec:conclusions}.

\section{The Runaway Velocities of Compact Objects in Unbound Binaries}
\label{sec:method}

Several studies have considered the effect of instantaneous mass loss on a binary's orbit \citep[e.g.,][]{blaauw1961, hills1970}, finding the landmark result that an initially circular binary will disrupt if it instantaneously loses at least half the system's total mass. \citet{hills1983} expanded upon this result by considering an initially eccentric binary and including the effects of a SN natal kick \citep[a separate derivation is also provided by][]{hurley2002}. \citet{hills1983} showed that if mass loss occurs near pericenter, then an eccentric binary could dissolve with substantially less mass lost, while near apocenter, a binary could withstand more mass lost. However, \citet{hills1983} and \citet{hurley2002} only described the properties of the binaries that remained bound. \citet{tauris1998} calculated the runaway velocities of binary components, but for initially circular binaries \citep[see also][]{gott1970}. The complete solution for the runaway velocities of stars disrupted from an initially eccentric binary, including the effect of mass loss and natal kick within the SN was separately provided by \citet{belczynski2008} and \citet{kiel2009} \citep[see also][]{pijloo2012}. For completeness, we reproduce that procedure, but in its non-dimensional form.

We consider a binary in which a helium star with mass $\Mi$ undergoes core collapse, evolving into a BH with mass $\Mf$, while the secondary star has a constant mass $M_2$. Although it is not important that the BH progenitor is a helium star for the sake of this derivation, the massive star progenitors of BHs are expected to be massive, helium-rich Wolf-Rayet stars \citep[for a review, see][]{langer2012}. It is useful to define two parameters dependent on the system's mass:
\begin{eqnarray}
\mu &=& \frac{\Mf M_2}{\Mf + M_2} \\
\beta &=& \frac{\Mf + M_2}{\Mi + M_2}.
\end{eqnarray}

At the instant of core collapse, the distance between the two components of a binary, $\bm{r}_0$, and its relative orbital velocity, $\bm{v}_0$, expressed in terms of the binary's orbital separation, $a$, and circular orbital velocity, $V_{\rm c} = \left[\mathcal{G}(\Mi+M_2)/a \right]^{1/2}$, respectively, are:
\begin{eqnarray}
\bm{r}_0 &=& \left[\cos E - e, \sqrt{1-e^2} \sin E, 0\right] \\
\bm{v}_0 &=& \frac{1}{1-e \cos E} \left[-\sin E, \sqrt{1-e^2} \cos E, 0\right],
\end{eqnarray}
where $e$ is the pre-SN orbital eccentricity and $E$ is the eccentric anomaly of the pre-SN orbit at core collapse. We have assumed the orbital plane lies in the $X-Y$ plane with pericenter aligned along the $X$ axis. \citet{belczynski2008} provide a schematic describing the orbital orientation (see their Figure 4). The individual stellar velocities around the center of mass are therefore:
\begin{eqnarray}
\bm{v}_1 &=& \frac{M_2}{\Mf + M_2} \bm{v_0} \\
\bm{v}_2 &=& \frac{\Mf}{\Mf + M_2} \bm{v_0}.
\end{eqnarray}
The helium star (subscript 1) receives a kick $\bm{v_k}$ (also expressed in terms of $V_{\rm c}$). 

The velocity of the post-SN center of mass is:
\begin{equation}
    \bm{v_{\rm CM}} = \frac{\Mf \left( \bm{v_1} + \bm{v_k} \right) + M_2\bm{v_2}}{\Mf + M_2},
\end{equation}
and the post-SN relative velocity of the two stars is:
\begin{equation}
    \bm{v} = \bm{v_1} - \bm{v_2} + \bm{v_k}.
\end{equation}

The specific angular momentum, $j = \left|\bm{j}\right|$, (in terms of $a$, $V_{\rm c}$, and $\mu$) is:
\begin{equation}
    \bm{j} = \bm{r_0} \times \bm{v}.
\end{equation}

The post-SN orbital energy can therefore be calculated:
\begin{equation}
    \mathcal{E} = \frac{v^2}{2} - \frac{\beta}{r_0},
\end{equation}
where $v = \left| \bm{v} \right|$ and $r_0 = \left| \bm{r_0} \right|$. In this work, we are only interested in systems that dissociate during core collapse, i.e., $\mathcal{E} > 0$. 

The post-SN hyperbolic orbit is defined by an eccentricity, $e$, and semi-latus rectum, $p$:
\begin{eqnarray}
    e &=& \sqrt{1 + 2 \mathcal{E} j^2 / \beta^2} \\
    p &=& j^2 / \beta.
\end{eqnarray}

Depending on the orbital parameters, the orbital phase where core-collapse occurs, and the kick velocity itself, the post-SN orbit will either be moving toward pericenter ($\bm{v}\cdot\bm{r_0} < 0$) or moving away from pericenter ($\bm{v}\cdot\bm{r_0} < 0$). We can define an angle, $\phi_0$, which measures how far along the hyperbola the two stars are immediately after core-collapse, as measured from the focus:
\begin{equation}
    \phi_0 = \begin{cases} 
            \cos^{-1} \left[ \left( p/r_0 - 1 \right) / e \right], \text{for } \bm{v}\cdot\bm{r_0} < 0 \\
            -\cos^{-1} \left[ \left( p/r_0 - 1 \right) / e \right], \text{for } \bm{v}\cdot\bm{r_0} > 0
    \end{cases}
\end{equation}
After some time, the binary's separation becomes arbitrarily large, and we can define the angle the BH leaves the system as
\begin{equation}
    \lim_{r_0\to\infty} \phi_f = \cos^{-1} \left( 1/e \right).
\end{equation}

The runaway velocity of the two stars can be calculated by performing a coordinate transformation, where we first apply a rotation $\mathcal{R}$ to align the post-SN angular momentum vector to $\hat{z}$ then a clockwise rotation $T$ to rotate the coordinate axis by $\phi_f - \phi_0$. In this coordinate frame, the final orbital velocity is:
\begin{equation}
    \bm{v}' = \sqrt{2 \mathcal{E}} T(\phi_f - \phi_0) \mathcal{R} \hat{r_0}.
\end{equation}

By applying the inverse transformation, we can now calculate the runaway velocities of the two components in our initial inertial frame:
\begin{eqnarray}
\bm{v_{\rm BH, f}} &=& -\frac{M_2}{\Mf + M_2} \mathcal{R}^{-1} (\bm{v}') + \bm{v_{\rm CM}} \\
\bm{v_{\rm 2, f}} &=& \frac{\Mf}{\Mf + M_2} \mathcal{R}^{-1} (\bm{v}') + \bm{v_{\rm CM}}.
\end{eqnarray}
We have verified the validity of this procedure by comparing it with direct orbital integration.

\section{Results}
\label{sec:results}

\begin{figure*}
    \begin{center}
    \includegraphics[width=1.0\textwidth]{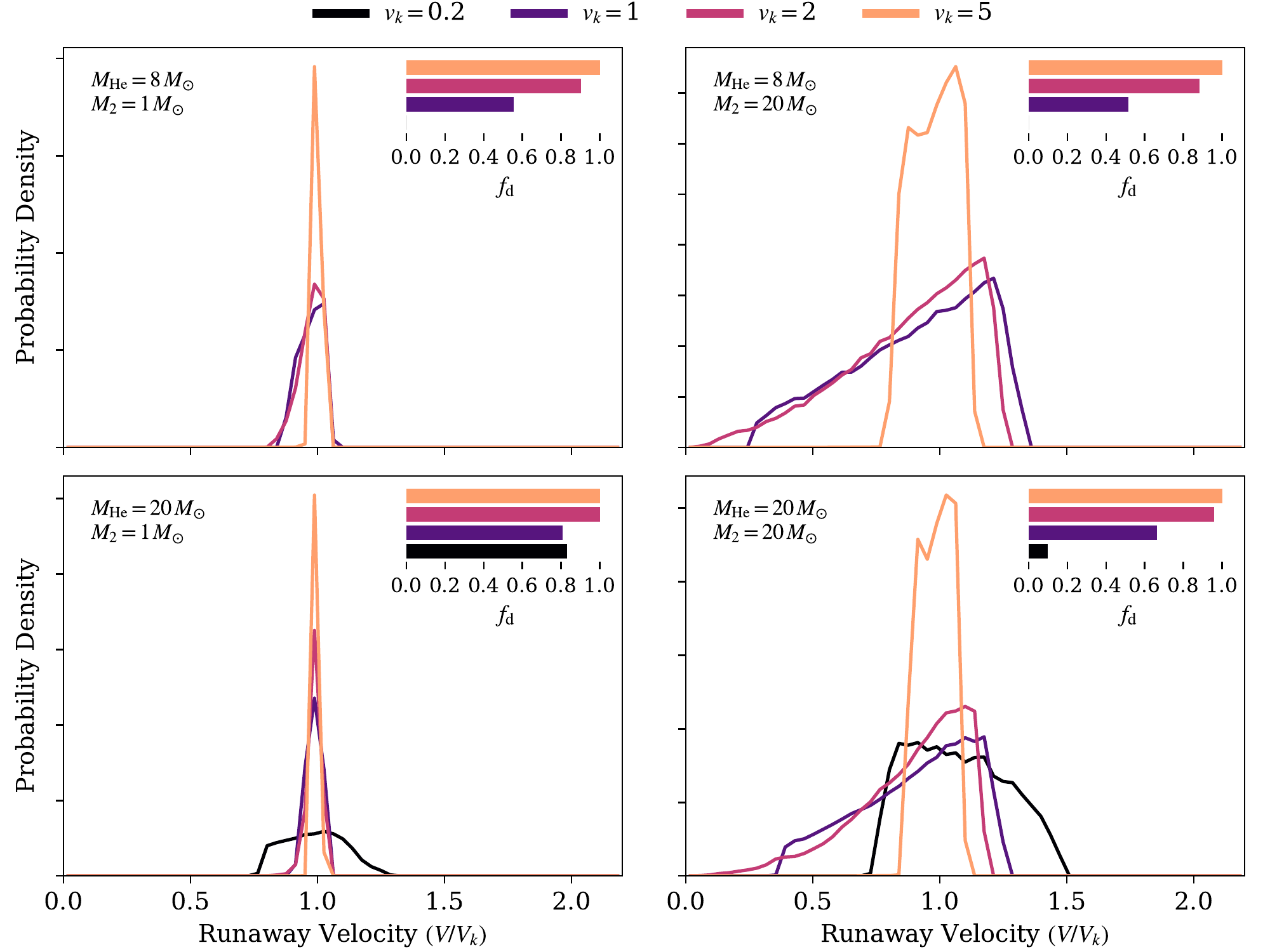}
    \caption{ Distributions of runaway velocities for possible \ogle{} progenitors for binaries with an initial eccentricity of 0.1, adopting a BH mass of 7.1\,\Msun{} from \citet{sahu2022}. Each panel shows a different combination of possible He-star progenitor masses and companion masses (left column shows low-mass companions and right column shows high-mass companions), while the curves in each panel show the effects of different kick velocities. In the top right of each panel we provide an inset displaying the fraction of systems that disrupt during core collapse. For most binary configurations, disruption velocities provide a close estimate of the kick velocity the BH received at birth. Binaries with more massive companions (right column of panels) have a somewhat broader distribution of runaway velocities. However, even in these cases, the runaway velocities provides an accurate estimate of the kick velocity. }  
    \label{fig:velocities_e01}
    \end{center}
\end{figure*}

\begin{figure*}
    \begin{center}
    \includegraphics[width=1.0\textwidth]{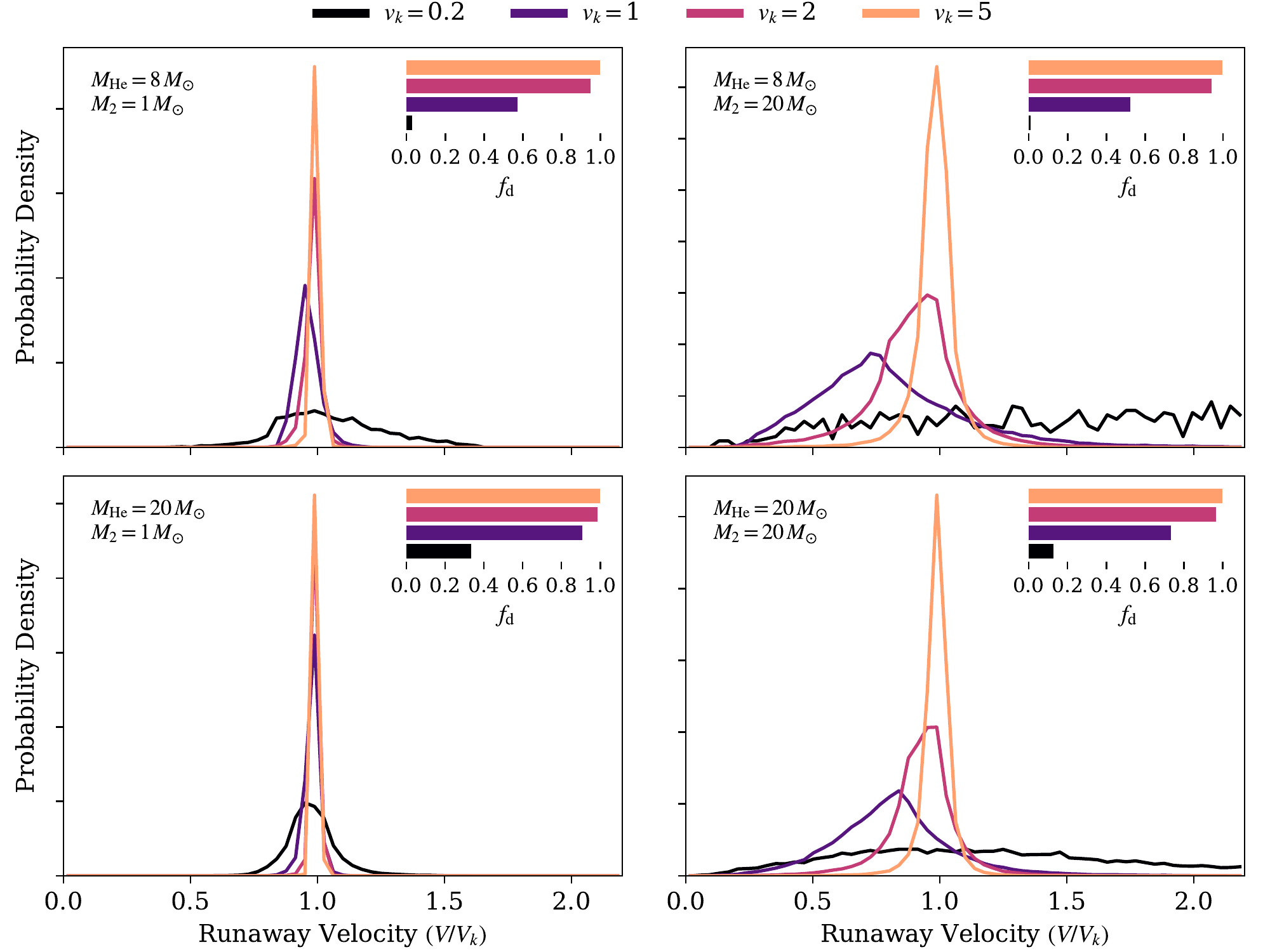}
    \caption{ We provide the runaway velocities for possible \ogle{} progenitors, for binaries with an initial eccentricity of 0.9, again adopting a BH mass of 7.1\,\Msun{}. Our results largely mimic those from Figure~\ref{fig:velocities_e01} except the distributions are even narrower, suggesting the runaway velocity is an even closer representation of the SN natal kick for highly eccentric pre-SN orbits. This is because eccentric binaries spends most of their time near apocenter, where the orbital velocities are small and the gravitational potential is shallow; runaway velocities are therefore dominated by the natal kicks received. Systems with high mass companions have distributions that peak toward somewhat lower runaway velocities, suggesting that the runaway velocities of these BHs may underestimate the natal kicks. Systems that are disrupted with lower kicks produce a much larger range of possible runaway velocities; however the insets in each panel suggest that these binaries rarely become disrupted during a SN. Only in the bottom, left panel, representing low-mass companions and high-mass He-star progenitors do low-kick systems tend to disrupt the binary with any frequency. The black curve in this panel demonstrates that the runaway velocity is an accurate representation of the BH's natal kick. }  
    \label{fig:velocities_e09}
    \end{center}
\end{figure*}

We use a Monte Carlo method to determine the effects of different SN kicks and mass loss on different possible pre-SN binary configurations. For each combination of He-star mass, companion mass, pre-SN orbital eccentricity, and kick velocity, we randomly and independently generate $10^5$ separate kick directions (drawn from an isotropic distribution across all $4\pi$ steradians) and eccentric anomalies (properly accounting for the fraction of an orbit an eccentric binary spends at each $E$). After processing each of our randomly drawn binaries using the method described in Section~\ref{sec:method}, we obtain a distribution of runaway velocities for those binaries that are disrupted during core-collapse. 

\begin{figure*}
    \begin{center}
    \includegraphics[width=1.0\textwidth]{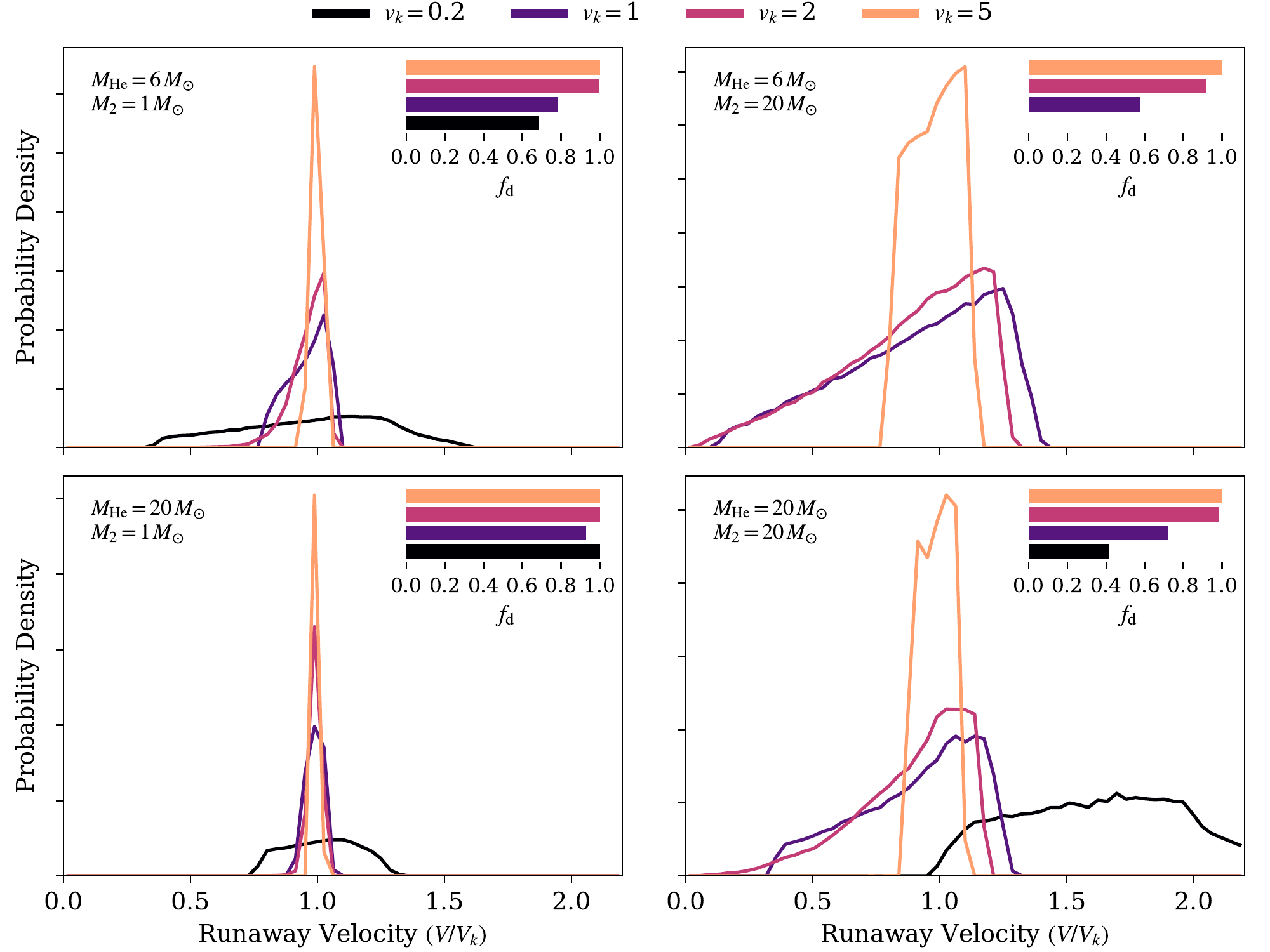}
    \caption{ We provide the runaway velocities for possible \ogle{} progenitors, for binaries with an initial eccentricity of 0.1, again adopting a BH mass of 2.1\,\Msun{} following \citet{lam2022}. Our results largely mimic those from Figures~\ref{fig:velocities_e01} and \ref{fig:velocities_e09}; the BH runaway velocities are an accurate proxy for the BH's natal kick. The only exception is in the bottom right panel, with large He-star masses ($\simeq$20\,\Msun) and large companion masses ($\simeq$20\,\Msun), in which case the BH's runaway velocity can be as much as twice the magnitude of its natal kick. Nevertheless, the black bar in the bottom right panel's inset shows that BHs receiving such relatively low kick velocities only have a $\simeq$40\% of disrupting. }  
    \label{fig:velocities_lam_e01}
    \end{center}
\end{figure*}

To explore the effects of different orbital parameters, we test somewhat extreme orbital configurations to demonstrate how these affect our results and with the idea that these will bracket the true progenitor to \ogle{}. We test both low-mass ($1\,\Msun$) and high-mass ($20\,\Msun$) companions, as well as both low-mass ($8\,\Msun$) and high-mass ($20\,\Msun$) He-star progenitors to the BH. To start, we adopt a mass of 7.1\,\Msun{} for \ogle{}, following \citet{sahu2022}. Figure~\ref{fig:velocities_e01} shows the resulting runaway velocities (as a fraction of the natal kick velocity) for these combinations of masses for different non-dimensional kick velocities of 0.2, 1, 2, and 5 (in terms of $V_c$) for a binary with an initial eccentricity of 0.1. 

The left column of panels shows that a low-mass companion results in runaway velocities within $\simeq$10\% of the kick velocity. The lack of a black curve in the top, left panel shows that when very little mass is lost from the orbit, a kick velocity similar to or greater than the orbital velocity is required to disrupt the binary. This is additionally expressed by the inset in each panel of Figure~\ref{fig:velocities_e01}, where we provide the fraction of systems disrupted during SN for each SN kick velocity. The inset in the bottom, left panel shows that when the He-star progenitor to the BH is sufficiently large ($\beta \lesssim$0.5), most binaries will disrupt, regardless of the natal kick received by the BH. These binaries all have runaway velocities very similar to the BH's natal kick. Even in the case of $v_k=0.2$, the runaway velocities are within 20\% of the BH natal kick velocity.

In the right column of panels in Figure~\ref{fig:velocities_e01} we show the runaway velocities for binaries with massive companions. These systems are more massive and therefore require a larger kick magnitude ($v_k\gtrsim$1) to disrupt with any notable frequency. The distribution of runaway velocities for these binaries peak near the BH natal kick velocity, albeit with a somewhat wider distribution compared with the left column of panels. The massive companion means that the He-star progenitor to the BH is initially orbiting with a much larger velocity (with respect to the system's center of mass) which leads to a wider distribution of possible runaway velocities, depending on the binary's orbital phase at core collapse and SN kick direction. Nevertheless, the runaway velocities tend to be within 20\% of the BH natal kick. For systems that receive moderate SN kicks ($v_k=$ 1 or 2), the distribution of possible runaway velocities extends to somewhat lower values, implying the SN kick could be somewhat larger than the observed runaway velocity.

\begin{figure*}
    \begin{center}
    \includegraphics[width=1.0\textwidth]{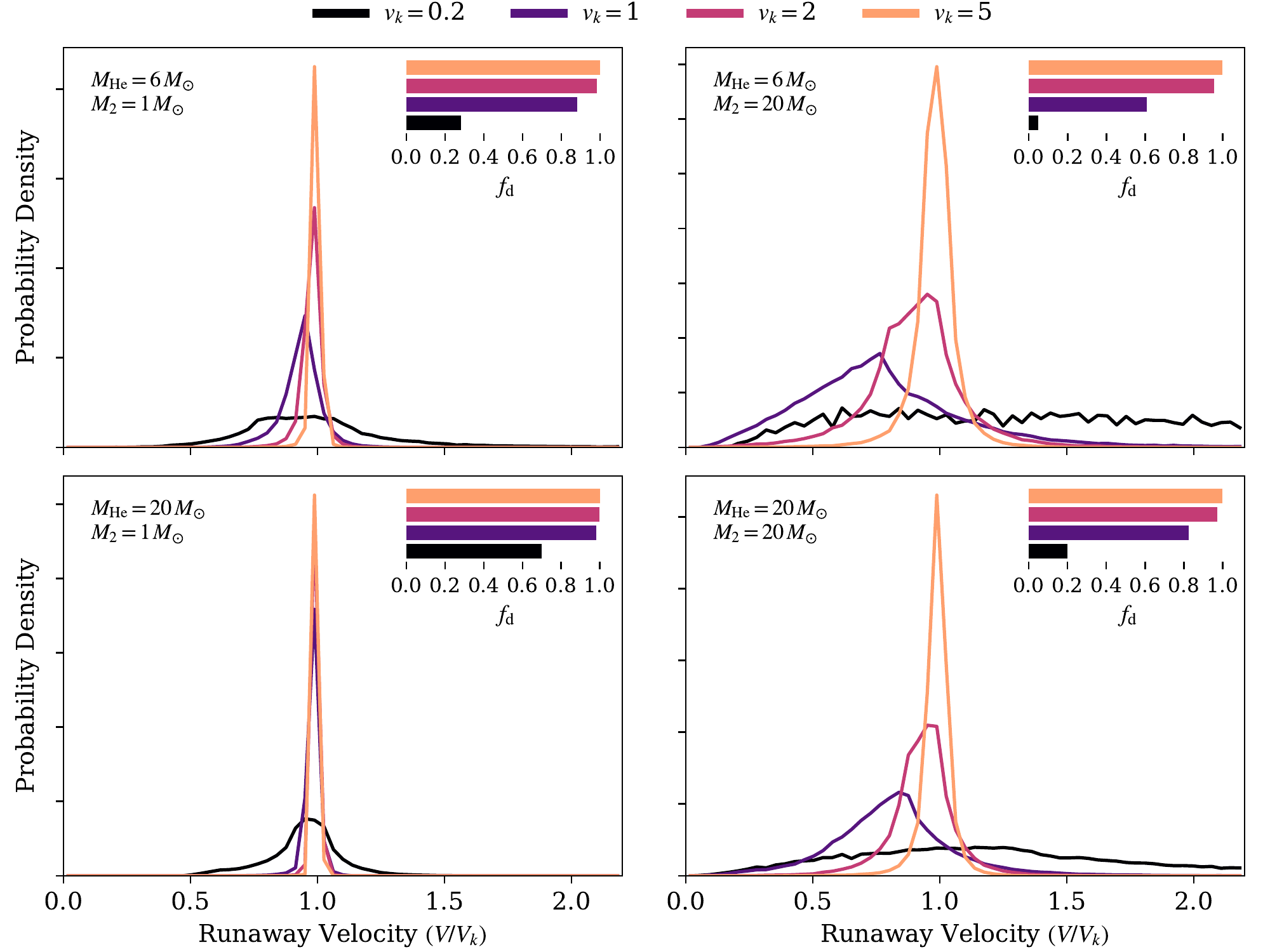}
    \caption{ We provide the runaway velocities for possible \ogle{} progenitors, for binaries with an initial eccentricity of 0.9, again adopting a BH mass of 2.1\,\Msun{} following \citet{lam2022}. As was the case for Figures~\ref{fig:velocities_e01}, \ref{fig:velocities_e09}, and \ref{fig:velocities_lam_e01}, the BH runaway velocities are an accurate proxy for the BH's natal kick. The only exceptions (those with low $v_k$ and large $M_2$) are unlikely to disrupt. }  
    \label{fig:velocities_lam_e09}
    \end{center}
\end{figure*}

In Figure \ref{fig:velocities_e09} we provide the same results as in Figure \ref{fig:velocities_e01}, but for binaries with an initial eccentricity of 0.9. Comparison between the two figures shows qualitatively similar results, with the largest difference being that the highly eccentric binaries have distributions of runaway velocities even more sharply peaked around the kick velocity. This is because highly eccentric binaries spend most of their orbits far from pericenter, where the orbital velocities and gravitational pull between the two components are low. In these cases, the SN kick plays the significantly dominant role in the outcome of the binary. A secondary effect can be seen in the $v_k = 1, 2$ curves in the right column of panels in Figure \ref{fig:velocities_e09}; these distributions are peaked at runaway velocities even smaller than the BH natal kick velocity, and extend to even lower velocities than that seen in Figure~\ref{fig:velocities_e01} for binaries with initial eccentricities of 0.1. If the true progenitor parameters of \ogle{} are well represented by these orbital configurations, then the runaway velocities in fact provide underestimates of the BH natal kick velocity.

For comparison, we additionally provide the distributions of runaway velocities adopting a BH mass of 2.1\,\Msun{} for \ogle, following the Solution 2 from \citet{lam2022}. Figure~\ref{fig:velocities_lam_e01} provides the resulting distributions for an eccentricity of 0.1 while Figure~\ref{fig:velocities_lam_e09} provides the same distributions but for an eccentricity of 0.9. We have adjusted the helium star mass, testing 6\,\Msun{} and 20\,\Msun{} models. In nearly all of the curves shown in Figures~\ref{fig:velocities_lam_e01} and \ref{fig:velocities_lam_e09} the runaway velocity is very similar to the BH's kick velocity, typically within 50\%, often much closer. As was the case with the distributions in Figures~\ref{fig:velocities_e01} and \ref{fig:velocities_e09} for a 7.1\,\Msun{} BH, the only exceptions occur for our $v_k=0.2$ model with $M_2=20$\,\Msun, in which case the runaway velocity can be as much as a factor of 2 different from $v_k$. However, such cases ought to be rare as the black bars in the insets of these panels indicate such binaries only have a $\lesssim$20\% probability of disrupting.

\begin{figure*}
    \begin{center}
    \includegraphics[width=1.0\textwidth]{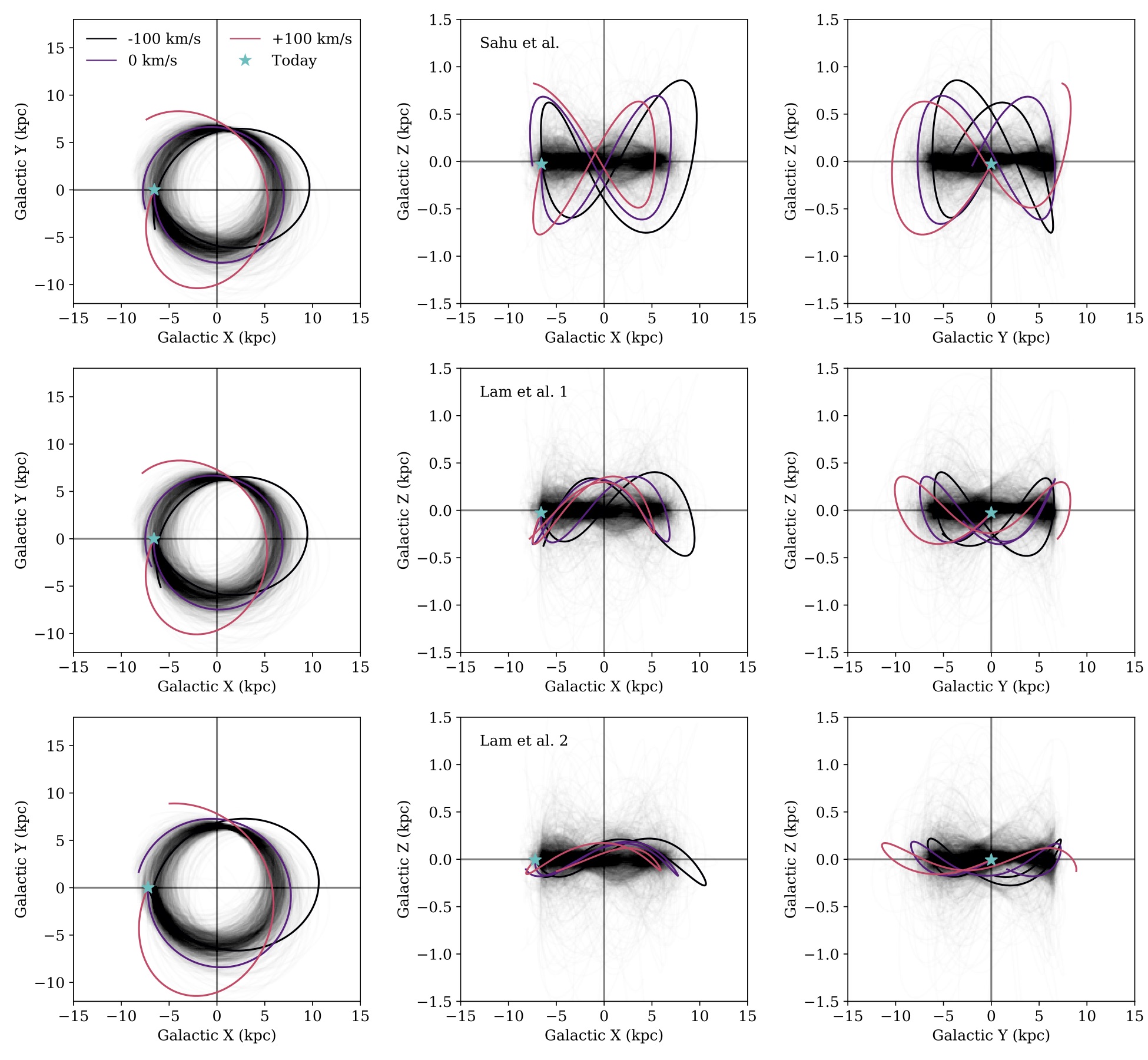}
    \caption{ The observed position, distance, and proper motion of \ogle{} provide five of the six dimensions of phase space describing its orbit. We calculate its orbit through the Milky Way adopting different possible radial velocities. Columns show different 2-dimensional projections of the resulting orbits, while rows show the solution from \citet{sahu2022} and the two solutions from \citet{lam2022}. The orbit is nearly circular, but can extend to a large height above the Milky Way disk. Black lines show Milky Way orbits of nearby field stars. \ogle's orbit is inconsistent with the thin disk, but is representative of the kinematic thick disk.}  
    \label{fig:galactic_orbit}
    \end{center}
\end{figure*}

The curves in Figures~\ref{fig:velocities_lam_e01} and \ref{fig:velocities_lam_e09} closely match those of Figures~\ref{fig:velocities_e01} and \ref{fig:velocities_e09}, calculated for a 7.1\,\Msun{} BH. The close similarities allow us to infer that our results can be applied to models with 3.7\,\Msun{} BHs following Solution 1 from \citet{lam2022} and indeed any reasonable mass for \ogle{}. Taken together, the results presented in Figures~\ref{fig:velocities_e01}, \ref{fig:velocities_e09}, \ref{fig:velocities_lam_e01}, and \ref{fig:velocities_lam_e09} lead us to conclude that the runaway velocity of BHs formed from a disrupting system provide an accurate proxy for the kicks these objects received at formation. While the correlation between the natal kick velocity and runaway velocity was pointed out by \citet{tauris1998} and \citet{kiel2009}, we show it to be robust for eccentric pre-SN orbits and highlight how it can actually be used to constrain natal kicks. {\it By measuring the peculiar velocity of a BH or other compact object that formed from a binary that disrupted during core collapse, we are in effect constraining the natal kick that object received.} However, as we discuss in the following sections, placing accurate constraints on the BH's natal kick requires knowledge of the pre-SN binary's peculiar velocity.

\section{\ogle's Peculiar Velocity}
\label{sec:peculiar_velocity}

In addition to the BH masses, the combined astrometric and photometric microlensing signals analyzed by \citet{sahu2022} and \citet{lam2022} provide distances and proper motions for \ogle{}. \citet{sahu2022} report a distance of $1.58\pm0.18$\,kpc and a proper motion of $\mu_{\alpha}=-4.36\pm$0.22\,mas\,yr$^{-1}$ and $\mu_{\delta}=3.06\pm$0.66\,mas\,yr$^{-1}$. \citet{lam2022} report two solutions. For Solution 1, \citet{lam2022} find a parallax of 0.65$^{+0.08}_{-0.08}$\,mas and a proper motion ($\mu_{\alpha}$, $\mu_{\delta}$) of ($-2.36^{+0.12}_{-0.13}$, 1.49$^{+0.73}_{-0.67}$)\,mas\,yr$^{-1}$. For Solution 2, \citet{lam2022} find a parallax of 1.09$^{+0.36}_{-0.33}$\,mas and a proper motion of ($-0.72^{+0.96}_{-0.94}$, 1.54$^{+1.29}_{-1.24}$)\,mas\,yr$^{-1}$. 

Since \citet{sahu2022} and \citet{lam2022} use the same photometric and astrometric data, the different astrometric solutions reported by these authors have their origin in the details of the lensing models they fit. In particular, both sets of authors report discrepancies between the photometric and astrometric lensing signal, with the photometric data preferring a heavier, faster moving compact object. Indeed, the two different solutions reported by \citet{lam2022} depend on how these authors comparatively weigh the photometric and astrometric data. Throughout this work, we consider all three solutions for \ogle{}.

\begin{figure}
    \begin{center}
    \includegraphics[width=1.0\columnwidth]{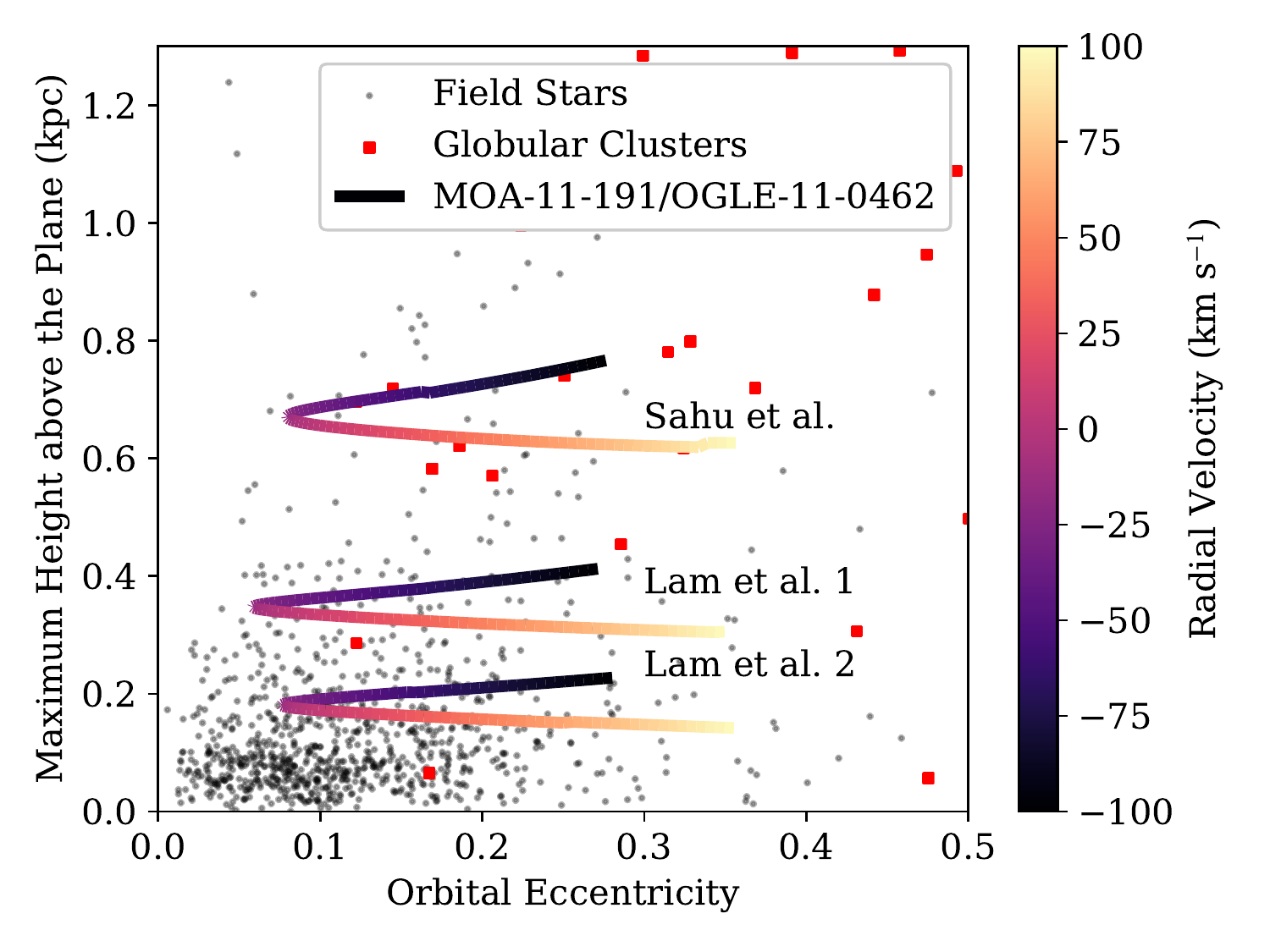}
    \caption{ The eccentricity of \ogle's orbit around the Milky Way with the maximum height it reaches above the Galactic Plane as a function of its radial velocity. For comparison we also show nearby field stars (black points) and globular clusters (red squares). Regardless of its radial velocity \ogle{} is consistent with being part of the kinematic thick disk.  }  
    \label{fig:e_zmax}
    \end{center}
\end{figure}

Along with the sky position, the derived distances (or parallaxes) and proper motions provide five of the six dimensions of phase space for \ogle{}; the radial velocity is unknown. In Figure~\ref{fig:galactic_orbit} we show the possible orbits in the Milky Way for \ogle{}, adopting possible radial velocities of $-$100, 0, and $+$100\,km\,s$^{-1}$ for each of the three possible solutions (different rows). The different columns show the orbits from face-on (left column) and edge-on from either side (middle and right columns). These orbits are produced by integrating \ogle{} backwards for 200 Myr using {\tt gala} \citep{gala} within the multi-component Milky Way model from \citet{bovy2015}, adopting {\tt astropy} v4.0 \citep[][]{astropy2013, astropy} constants for the position and peculiar velocity of the Sun. For comparison in Figure~\ref{fig:galactic_orbit} we additionally show the orbits (black lines) for stars in the same vicinity of \ogle{}, identified by selecting all Gaia EDR3 \citep{Gaia_mission, gaia_edr3, lindegren2021} stars within 2$^{\circ}$ of \ogle{} that have a parallax divided by the parallax uncertainty of at least 10, a parallax between 0.55 and 0.7 mas, and radial velocities measured by Gaia DR2 \citep{Gaia_radial_velocities}.

The first row of Figure~\ref{fig:galactic_orbit}, shows that variations in radial velocity by $\pm$100\,km\,s$^{-1}$ largely effect the eccentricity of the Galactic orbit, but do not strongly affect the maximum height above or below the Milky Way disk, $Z_{\rm max}$. This is a result of the strategy that microlensing surveys typically focus on the Galactic Bulge where stellar densities are high to maximize the likelihood of observing an event \citep{mroz2019}. This perspective allows for a lens's motion in the Galactic $Z$ direction to be well measured. Fortunately, $Z_{\rm max}$ is also a reasonable indicator of the Milky Way population to which a star belongs. All the orbits for \ogle{} in Figure~\ref{fig:galactic_orbit} are consistent with disk orbits, regardless of which astrometric solution or radial velocity is adopted (excluding extreme velocities with magnitudes $>>$100\,km\,s$^{-1}$). Comparison against stars in the same vicinity shows that the $Z_{\rm max}$ value is sensitive to which astrometric solution is adopted.

In Figure~\ref{fig:e_zmax} we compare the orbital eccentricities and $Z_{\rm max}$ of nearby stars against \ogle{} for each of the three astrometric solutions, adopting a range of possible radial velocities. This figure demonstrates that the kinematic thin disk is clearly separate from any of the three solutions for \ogle{}, although Solution 2 from \citet{lam2022} is close. However, \ogle{} is consistent with orbits from the Milky Way thick disk. There are two possibilities: either \ogle{} was born in the kinematic thick disk or it was formed in the kinematic thin disk and was kicked into a thick-disk-like orbit. In the former, constraints are difficult to place as thick disk stars span a wide range of kinematics characteristics; the different radial velocity solutions for \ogle{} in Figure~\ref{fig:e_zmax} show that a thick disk star could receive a kick as large as 100\,km\,s$^{-1}$ and still appear as a thick disk star. 

\begin{figure}
    \begin{center}
    \includegraphics[width=1.0\columnwidth]{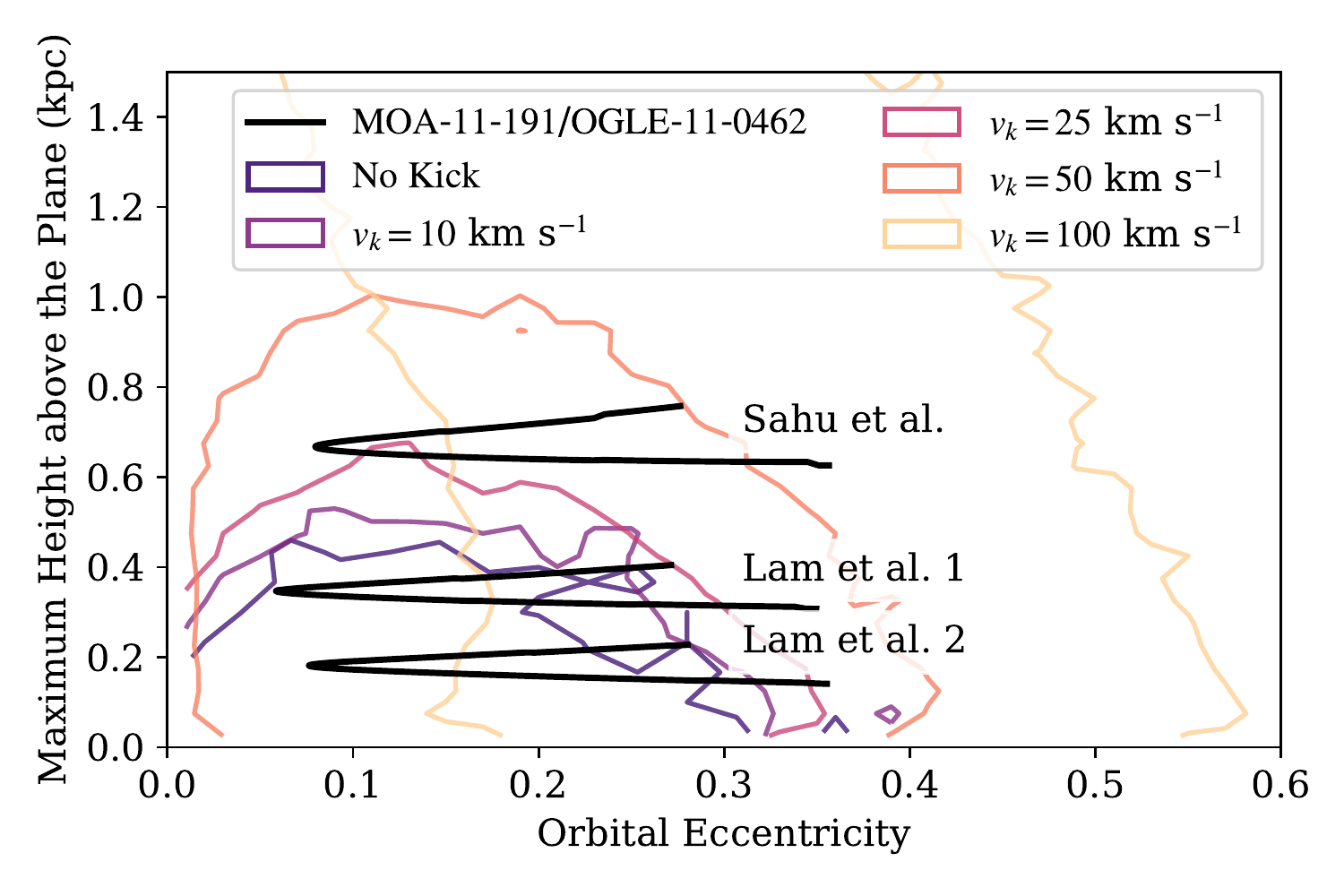}
    \caption{ We compare orbital eccentricity (in the Milky Way) and maximum height reached above the Galactic Plane for \ogle{} and for populations of nearby stars after artificially applying different velocity kicks to their trajectories. Contours enclose 90\% of the samples, and are therefore representative of the kinematic thin disk. If \ogle{} were formed in the thin disk and subsequently kicked out, it likely had to receive a kick velocity in excess of 50 km s$^{-1}$. }  
    \label{fig:e_zmax_kick}
    \end{center}
\end{figure}

Because the kinematic thin disk is dynamically colder, we can place constraints on the potential kick \ogle{} received at birth. To each star in our sample of stars in the vicinity of \ogle{}, we apply 10$^3$ random kicks, isotropically distributed with magnitudes of 10, 25, 50, and 100\,km\,s$^{-1}$, then integrate their orbits to find their orbital eccentricity and $Z_{\rm max}$. Based on our results described in Section~\ref{sec:results}, these kicks correspond identically to the natal kicks of single BHs as well as those of BHs formed out of the disruption of a stellar binary. Contours in Figure~\ref{fig:e_zmax_kick} contain the 90\% confidence interval of the resulting distributions. Increasing kick velocities lead to both an increase in the orbital eccentricity as well as $Z_{\rm max}$. Different constraints can be placed on the peculiar velocity of \ogle{}, depending on which astrometric solution is adopted. For the solution from \citet{sahu2022}, we find a kick velocity of 50-100\,km\,s$^{-1}$ is required. Alternatively, Solutions 1 and 2 from \citet{lam2022} allow kick velocities of up to $\simeq$50\,km\,s$^{-1}$, with Solution 2 more consistent with lower kick velocities. We stress that these constraints are statistical, based on a qualitative comparison between \ogle{} and nearby stars in its vicinity for only two parameters summarizing their orbit in the Galaxy. Furthermore, they are under the assumption that \ogle{} formed within the kinematic thin disk and has a radial velocity within $\simeq$100 km s$^{-1}$ of zero. In the case that \ogle{}'s unknown radial velocity is significantly non-zero (which is very unlikely based on viewing angle statistics), the required kick can become arbitrarily large.

\section{How did \ogle{} Form?}
\label{sec:discussion}

Here we consider multiple scenarios for how \ogle{} could have formed.

\subsection{Single Star Scenario}

Although $>$70\% of massive stars have binary companions \citep{sana2012} and most isolated BHs have a binary origin \citep{vigna-gomez2022}, we will first consider the possibility that \ogle{} was one of the minority of massive stars that formed and evolved as a single star. Given its Galactic orbit described in Figures~\ref{fig:galactic_orbit} and \ref{fig:e_zmax}, \ogle{} either formed in the kinematic thick disk and received a kick $\lesssim$100\,km\,s$^{-1}$ or was formed in the thin disk and received a kick of at most a few tens of km\,s$^{-1}$ into its current orbit, with a value dependent upon which astrometric solution is adopted. Since it is a BH, we have no constraints on \ogle{}'s age; BHs of a few \Msun{} can be formed at any metallicity \citep[e.g.,][]{riley2022}. The relatively likelihood of whether \ogle{} was formed in the thick disk or kicked out of the thin disk is determined only from an astrophysical prior based on the relative mass of the two populations and probability that they produce a detectable microlensing signal. However, we note that in the single star scenario, any perturbation to \ogle{}'s kinematics is due to the kick it received during core collapse; no other process in the evolution of isolated single stars can produce substantial kicks. We therefore conclude that, if \ogle{} was formed as a single star, it likely received a natal kick $\lesssim$100\,km\,s$^{-1}$.

\begin{figure}
    \begin{center}
    \includegraphics[width=1.0\columnwidth]{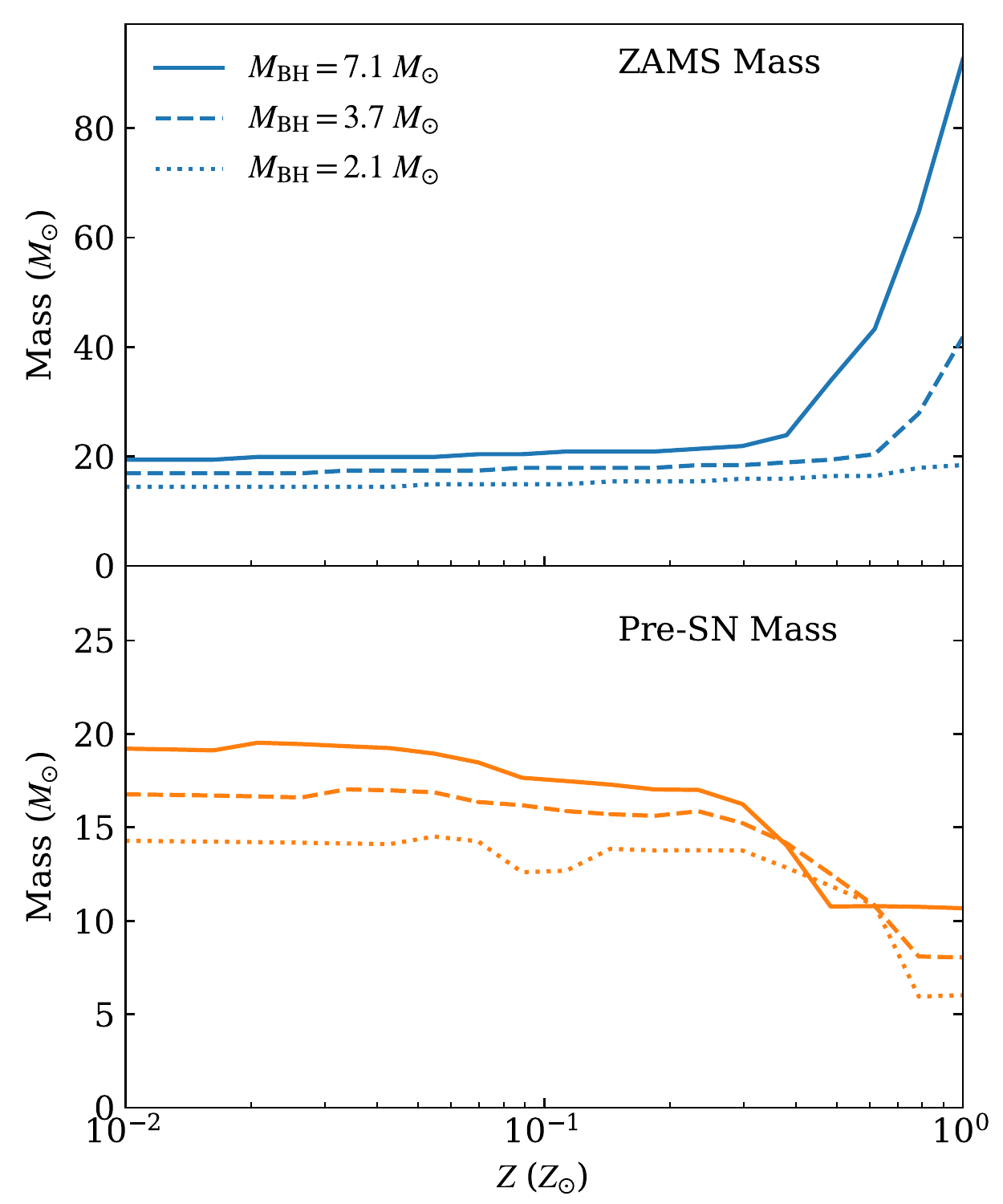}
    \caption{ We show the initial masses and pre-SN masses of single stars that produce 2.1, 3.7, and 7.1\,\Msun\ BHs using \cosmic. The possible progenitors for \ogle{} depend strongly on metallicity: The ZAMS masses are within a broad range of 15 to 90\,\Msun, while the pre-SN masses are constrained to be between 6 and 20\,\Msun. Binary interactions could alter this range somewhat. }  
    \label{fig:pre-SN_masses}
    \end{center}
\end{figure}

\subsection{Ejection from a Globular Cluster}

The dynamics at the centers of globular clusters, especially those that are massive and core-collapsed, can easily expel BHs into the Milky Way field. Typical dynamical encounters that expel stars will cause them to only barely escape the system, resulting in orbits that are similar to its host globular cluster. In Figure~\ref{fig:e_zmax} we provide the orbital eccentricities and $Z_{\rm max}$ of known globular clusters from the catalog of \citet{baumgardt2021} and \citet{vasiliev2021}, using updated Gaia EDR3 astrometry \citep{lindegren2021, gaia_edr3}. Although some globular clusters have orbits consistent with \ogle{}, particularly the solution from \citet{sahu2022}, most have halo-like orbits. We therefore consider a globular cluster origin for \ogle{} unlikely.

\subsection{Formation From a Disrupted Binary}

We consider the possibility that \ogle{} formed initially in a binary system that was disrupted during BH formation. One possibility is that the core collapse event forming \ogle{} disrupted the binary system. In Figure~\ref{fig:pre-SN_masses} we use the binary population synthesis code {\tt COSMIC} with default code options \citep[see][and references therein]{breivik2020} to determine initial masses and pre-SN masses of stars that evolve into 2.1, 3.7, and 7.1\,\Msun\ BHs. Although metallicity has a strong effect on the stars that could evolve into \ogle{}, the pre-SN masses (bottom panel) are constrained to be within 6 and 20\,\Msun. Even though binary interactions could lower this mass range somewhat, Figure~\ref{fig:pre-SN_masses} suggests that the runaway velocity distributions in Figures~\ref{fig:velocities_e01} and \ref{fig:velocities_e09} are applicable to the formation of \ogle. Under this scenario, \ogle{} was unlikely to have formed with a kick velocity $\gtrsim$100\,km\,s$^{-1}$ regardless of whether it formed in the kinematic thin or thick disk.

An alternative possibility is \ogle{}'s progenitor was ejected from a binary system when its more massive companion collapsed, and \ogle{} itself collapsed into a BH somewhat later. From the top panel of Figure~\ref{fig:pre-SN_masses} we can derive possible binary configurations. At $Z_{\odot}$, the primary in the system would have a pre-SN mass $\gtrsim$10\,\Msun{}, while \ogle{}'s progenitor would have been $\gtrsim$40\,\Msun. In this scenario, \ogle{}'s progenitor is the more massive of the two stars, suggesting its orbital velocity compared to the system's center of mass \citep[and therefore its runaway velocity upon the binary's disruption;][]{tauris1998} will be small. \ogle{}'s peculiar velocity is therefore likely to be dominated by whatever kick it received upon its own core collapse. At lower metallicity the primary star could have been more massive at its core collapse, and \ogle{}'s progenitor could have received a significant kick were the binary in a relatively tight orbit upon disruption. \ogle{} would then receive an additional kick when its own BH collapsed somewhat later. 

One last possibility is that \ogle{} was the secondary in a binary that survived its first SN, but was disrupted in the second SN. Although two SNe are required to form double compact objects, it is unlikely that \ogle{} formed from the second SN in such a system. \citet{renzo2019} find $\simeq$86\% of massive binaries disrupt during the primary's core collapse. The surviving systems must avoid a merger during subsequent mass transfer sequences to remain bound by the time the secondary undergoes core collapse \citep{gallegos-garcia2021}. We therefore focus on scenarios in which the system formed from the disruption of the first SN in a binary.

\section{Discussion and Conclusions}
\label{sec:conclusions}

We calculate the runaway velocities of compact objects ejected from disrupted binaries. While our analysis focuses on BHs, as we are primarily interested in its implications for BH microlenses, this approach can also be applied to runaway neutron stars, a topic already the subject of numerous studies \citep{gott1970, radhakrishnan1985, tauris1998}. While the connection between a compact object's natal kick and its runaway velocity was made by both \citet{tauris1998} and \citet{kiel2009}, it was not recognized as a way to constrain natal kicks. \citet{tauris1998} showed that the runaway velocity is principally dependent upon the natal kick velocity, while other parameters such as the component masses have little effect. By using Monte Carlo random draws \citet{kiel2009} show that the distribution of resultant runaway velocities are similar to the input Maxwellian distribution of natal kick velocities. In this work, we focus on individual systems, showing that the runaway velocity is typically within $\simeq$20\% of the compact object's natal kick. 

Of course, for sufficiently tight orbits or small natal kicks such that the kick velocity is a small fraction of the orbital velocity, the correlation between runaway velocity and natal kick velocity decreases. Nevertheless, in our Monte Carlo simulations, even small kicks ($v_{\rm k}=0.2$) demonstrate a significant correlation. How is this possible? Kicks significantly smaller than the orbital velocity typically do not disrupt the binary unless a substantial amount of mass is lost from the binary. For eccentric binaries that spend most of their orbits far from periastron, more than half of the system's mass must be lost to disrupt \citep{hills1983}. At the same time, the orbital velocities far from periastron are lower than their circular velocity equivalent. Therefore, systems with initially more eccentric orbits will produce the tightest correlation between runaway velocities and natal kicks.

To determine the eccentricity distribution of pre-SN orbits, binary population synthesis including an accurate distribution of binaries at the zero-age main sequence is required. For sufficiently close orbits, mass transfer and tidal interactions ought to circularize binaries prior to core collapse \citep{van_den_heuvel1976}. Yet if we are interested in the distribution of runaway BHs, orbits of all periods, including those with wide orbits that may never circularize prior to core collapse, ought to be considered. The tests we run with different mass combinations, kick velocities, and eccentricities suggest that our main result holds over a wide range of orbits: measuring the peculiar velocity of a compact object provides a tight constraint of the kick velocity it received at birth.

We apply this result to the recently detected BH \ogle{}, which has three solutions for mass, distance, and proper motion, each of which we separately analyze. Although the lens's radial velocity is unmeasured, we find that for reasonable values, \ogle's motion through the Milky Way is consistent with being part of the kinematic thick disk. Alternatively were its progenitor formed in the kinematic thin disk, our analysis suggests \ogle{} received a kick $\lesssim$100 km s$^{-1}$, depending on which solution is adopted.

While the natal kick constraints we place on \ogle{} are, perhaps, quite broad, the method highlights the potential for constraints derived from future, individual well-observed systems or even populations. Recently, \citet{vigna-gomez2022} use binary population synthesis to analyze the formation of \ogle{} and suggest it -- and other BH microlenses like it -- can be used to constrain binary evolution parameters in general. In this work, we focus only on BH natal kicks, arguing that they are especially constrained by astrometric microlenses with well-measured proper motions. 

When considering future observations, two competing effects lead to detection biases that ought to be accounted for. Faster moving BHs are more likely to be detected, as the likelihood of lensing a background star scales with its tangential velocity. At the same time, degeneracy in photometric microlensing solutions implies that lenses with longer timescales (and are therefore likely to be massive) are more likely to be followed up with astrometric observations. Regardless, the addition of future astrometric BHs with precise proper motion and parallax measurements will only improve the constraints on BH natal kicks.

\acknowledgements

We thank the anonymous referee for a suggestions which improved the quality of this manuscript.
We thank Jessica Lu and Casey Lam for useful discussions.
J.J.A.\ acknowledges support from CIERA and Northwestern University through a Postdoctoral Fellowship. V.K.\ was partially supported through a CIFAR Senior Fellowship and a Guggenheim Fellowship. 
This work has made use of data from the European Space Agency (ESA) mission {\it Gaia} (\url{https://www.cosmos.esa.int/gaia}), processed by the {\it Gaia} Data Processing and Analysis Consortium (DPAC,
\url{https://www.cosmos.esa.int/web/gaia/dpac/consortium}). Funding for the DPAC has been provided by national institutions, in particular the institutions participating in the {\it Gaia} Multilateral Agreement.

\software{{\tt astropy} \citep{astropy}, {\tt gala} \citep{gala}, {\tt matplotlib} \citep{matplotlib}, {\tt NumPy} \citep{numpy}, {\tt SciPy} \citep{scipy}}

\bibliographystyle{aasjournal}
\bibliography{gaia}

\end{document}